\documentclass[12pt]{article}
\usepackage{axodraw}
\usepackage{epsfig}
\def\e{{\rm e}}
\newcommand{\be}{\begin{equation}}
\newcommand{\ee}{\end{equation}}
\newcommand{\bea}{\begin{eqnarray}}
\newcommand{\eea}{\end{eqnarray}}

\newcommand{\gm}{\gamma}
\newcommand{\Gm}{\Gamma}

\newcommand{\ep}{\epsilon}

\newcommand{\dd}{\mbox{d}}

\newcommand{\lra}{\leftrightarrow}

\newcommand{\nn}{\nonumber}

\begin{document}
\parindent=1.5pc

\begin{titlepage}
\rightline{hep-ph/0209193}
\rightline{September 2002}
\bigskip
\begin{center}
{{\bf The Leading Power Regge Asymptotic Behaviour of
Dimensionally Regularized Massless
On-Shell Planar Triple Box
}} \\
\vglue 5pt
\vglue 1.0cm
{ {\large V.A. Smirnov\footnote{E-mail: smirnov@theory.sinp.msu.ru}
} }\\
\baselineskip=14pt
\vspace{2mm}
{\em Nuclear Physics Institute of Moscow State University}\\
{\em Moscow 119899, Russia}
\vglue 0.8cm
{Abstract}
\end{center}
\vglue 0.3cm
{\rightskip=3pc
 \leftskip=3pc
\noindent
The leading power asymptotic behaviour
of the dimensionally regularized massless
on-shell planar triple box diagram in the Regge limit $t/s \to 0$
is analytically evaluated.
\vglue 0.8cm}
\end{titlepage}

Systematical analytical evaluation of two-loop Feynman diagrams with
four external lines within dimensional regularization \cite{dimreg}
began three years ago.
In the pure massless case with all end-points on-shell, i.e.
$p_i^2=0,\;i=1,2,3,4$, the problem of analytical evaluation
of two-loop four-point diagrams in expansion in $\ep=(4-d)/2$,
where $d$ is the space-time dimension, has been completely solved
in \cite{K1,SV,Tausk,AGO,ATT,GR3}.
The corresponding analytical algorithms have been successfully
applied to the evaluation of two-loop virtual
corrections to various scattering processes \cite{appl} in the
zero-mass approximation.

In the case of massless two-loop four-point diagrams with
one leg off-shell the problem of the evaluation has been solved
in \cite{S2,GR2}, with subsequent applications \cite{appl3j} to the
process $e^+e^-\to 3$jets.
(See \cite{MGU} for recent reviews of the present status
of NNLO calculations. See \cite{S4} for a brief review of results on the
analytical evaluation of various double-box Feynman integrals
and the corresponding methods of evaluation.) For
another three-scale calculational problem, where
all four legs are on-shell and there is
a non-zero internal mass, a first analytical result was obtained
in \cite{S3} for the scalar master double box.

The purpose of this paper is to turn attention to three-loop
on-shell massless four-point diagrams. As a first step,
the leading power asymptotic behaviour
of the dimensionally regularized massless
on-shell planar triple box diagram shown in Fig.~1
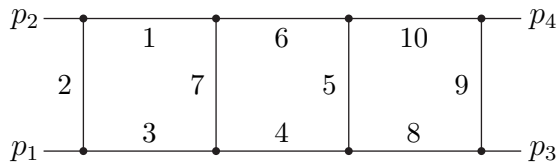
\begin {figure} [htbp]
\begin{picture}(400,80)(-140,-10)
\Line(-15,0)(0,0)
\Line(-15,50)(0,50)
\Line(165,0)(150,0)
\Line(150,0)(100,0)
\Line(150,50)(100,50)
\Line(165,50)(150,50)
\Line(0,0)(50,0)
\Line(50,0)(100,0)
\Line(100,50)(50,50)
\Line(50,50)(0,50)
\Line(0,50)(0,0)
\Line(50,0)(50,50)
\Line(100,0)(100,50)
\Line(150,0)(150,50)
\Vertex(0,0){1.5}
\Vertex(50,0){1.5}
\Vertex(100,0){1.5}
\Vertex(0,50){1.5}
\Vertex(50,50){1.5}
\Vertex(100,50){1.5}
\Vertex(150,0){1.5}
\Vertex(150,50){1.5}
\Text(-22,0)[]{$p_1$}
\Text(174,0)[]{$p_3$}
\Text(-22,50)[]{$p_2$}
\Text(174,50)[]{$p_4$}
\Text(25,43)[]{\small 1}
\Text(-7,25)[]{\small 2}
\Text(25,7)[]{\small 3}
\Text(75,7)[]{\small 4}
\Text(43,25)[]{\small 7}
\Text(93,25)[]{\small 5}
\Text(75,43)[]{\small 6}
\Text(125,7)[]{\small 8}
\Text(125,43)[]{\small 10}
\Text(143,25)[]{\small 9}
\end{picture}
\vspace*{2mm}\\
\caption{Planar triple box diagram.}
\end{figure}
%
in the Regge limit $t/s \to 0$ will be analytically evaluated.
This calculation will demonstrate that a three-loop BFKL analysis
\cite{BFKL} (at least its virtual part which can be reduced to
the evaluation of Regge asymptotics) is
possible\footnote{In three loops, non-planar diagrams
as well as higher terms of expansion of double boxes in $\ep$
are also needed.}.

The calculation will be based on the technique of
alpha parameters and Mellin--Barnes (MB) representation
which was successfully used in \cite{K1,Tausk,S2,S3} and reduces,
due to taking residues and shifting contours, to
a decomposition of a given MB integral into pieces
where a Laurent expansion of the integrand in $\ep$ becomes possible.
At a final stage, summation formulae for  series of
$S_1 (n) S_3 (n)/n^2$, $\psi''(n+1) S_2(n)/n$
etc., where
$S_k(n) =  \sum_{j=1}^n j^{-k}$, are used.
A table of such formulae is presented in Appendix.

The general planar triple box Feynman integral without numerator
takes the form
\bea
T(a_1,\ldots,a_{10};s,t;\ep) &=&
\int\int\int \frac{\dd^dk \, \dd^dl \, \dd^dr}{(k^2)^{a_1}
[(k+p_2)^2]^{a_2}
[(k+p_1+p_2)^2]^{a_3}}
\nn \\ && \hspace*{-10mm}
\times \frac{1}{
[(l+p_1+p_2)^2]^{a_4}[(r-l)^2]^{a_5}
(l^2)^{a_6} [(k-l)^2]^{a_7} }
\nn \\ && \hspace*{-10mm}
\times \frac{1}{
[(r+p_1+p_2)^2]^{a_8} [(r+p_1+p_2+p_3)^2]^{a_9}
(r^2)^{a_{10}} }
\, ,
\label{3box}
\eea
where $s=(p_1+p_2)^2$ and $t=(p_2+p_3)^2$ are Mandelstam variables, and
$k,l$ and $r$ are loop momenta.
Usual prescriptions $k^2=k^2+i 0, \; s=s+i 0$, etc. are implied.

To evaluate the leading power asymptotic behaviour of the master triple
box (\ref{3box}), i.e. for all $a_i=1$, in the limit $t/s \to 0$ one
can use the strategy of expansion by regions \cite{BS,SR,Sb}.
It shows that
in the leading power only (1c-1c-1c) and (2c-2c-2c) regions contribute,
with the leading power behaviour $1/t$.
(See \cite{SR} and Chapter~8 of \cite{Sb} for definitions of these
contributions.)
The leading power (2c-2c-2c) contribution for the general planar triple box
takes the form
\bea
T^{(2c-2c-2c)}(a_1,\ldots,a_{10};s,t;\ep)
 &=&
\int\int\int \frac{\dd^dk \, \dd^dl \, \dd^dr}{(k^2)^{a_1}[(k+p_2)^2]^{a_2}
(2 p_1 k+s)^{a_3}}
\nn \\ && \hspace*{-58mm}
\times \frac{1 
}{
(2 p_1 l+s)^{a_4}[(r-l)^2]^{a_5}
(l^2)^{a_6} [(k-l)^2]^{a_7} (2p_1 r+s)^{a_8} [(r+p_2+\tilde{p})^2]^{a_9}
(r^2)^{a_{10}}  }\;,
\label{3box-c}
\eea
where $\tilde{p}$ is such that
$\tilde{p}^2=t, \; 2p_1 \tilde{p}=0, \; 2p_2 \tilde{p}=-t$.
The leading power (1c-1c-1c) contribution is obtained due to the
symmetry $\{1\lra 3,\;4\lra 6,\;8\lra 10\}$.

To resolve the singularity structure of Feynman integrals in $\ep$
it is very useful to apply the MB representation
\be
\frac{1}{(X+Y)^{\nu}} = \frac{1}{\Gm(\nu)}
\frac{1}{2\pi i}\int_{-i \infty}^{+i \infty} \dd z
\frac{Y^z}{X^{\nu+z}} \Gm(\nu+z) \Gm(-z) \;,
\label{MB}
\ee
that makes it
possible to  replace sums of terms raised to some power by their
products in some powers, at the cost of introducing extra
integrations.
In \cite{K1,Tausk,S2}
MB integrations were introduced directly in alpha/Feynman
parametric integrals.
It turns out more convenient
to follow (as in \cite{ATT,S3}) the strategy of \cite{UD} and introduce,
in a suitable way,
MB integrations, first, after integration over one of the loop
momenta, $r$, then after integration over $l$,
and complete this procedure after integration
over the loop momentum, $k$.

After appropriate changes of variables
we arrive at the following  fourfold MB representation of
(\ref{3box-c}):
\bea
T^{(2c-2c-2c)}(a_1,\ldots,a_{10};s,t;\ep) &=&
\frac{\left(i\pi^{d/2} \right)^3 (-1)^a}{
\Gm(4-a_{589(10)}-2\ep)
(-s)^{a_{348}} (-t)^{a_{125679(10)}-6+3\ep}}
\nn \\ &&  \hspace*{-60mm}\times
\frac{\Gm(6 - a_{15679(10)}- 3 \ep)
\Gm(a_{125679(10)}-6 + 3 \ep)}{\prod_{j=2,5,7,8,9,10}\Gm(a_j)}
\frac{1}{(2\pi i)^4} \int_{-i\infty}^{+i\infty}
\prod_{j=1}^4 \dd z_j \;
\Gm(a_8 + z_3)\Gm(-z_3)
\nn \\ &&  \hspace*{-60mm}\times
\frac{\Gm(4 - a_{569(10)}- 2 \ep + z_1)\Gm(-z_1)
\Gm(a_{5679(10)}-4  + 2 \ep + z_2)
\Gm(6 - a_{125679(10)}- 3 \ep - z_2)}
{\Gm(8 - a_{1235679(10)}- 4 \ep + z_1)
\Gm(a_{15679(10)}-4 + 2 \ep + z_2)
\Gm(a_4 - z_3)}
\nn \\ &&  \hspace*{-60mm}\times
\frac{
\Gm(a_{15679(10)}- a_3 -4 + 2 \ep + z_1 + z_2)
\Gm(a_4 + z_1 - z_3)
\Gm(2  - a_{9(10)}- \ep + z_3) \Gm(z_2 - z_4) }
{\Gm(6 - a_{45679(10)}- 3 \ep + z_3)
\Gm(a_{569(10)}-2 + \ep + z_4)}
\nn \\ &&  \hspace*{-60mm}\times
\Gm(a_9 + z_4)\Gm(a_{59(10)} - 2 + \ep + z_4)
\nn \\ &&  \hspace*{-60mm}\times
\Gm(2 - a_{47} - \ep - z_1 - z_2 + z_3 + z_4)
\Gm(2 - a_{589} - \ep - z_3 - z_4)
\, ,
\label{2ccc-MB}
\eea
where $a=\sum_{i=1}^{10} a_i$, $a_{589(10)}=a_5+a_8+a_9+a_{10},
a_{348}=a_3+a_4+a_8$, etc., and
integration contours are chosen in the standard way.

We then turn to the master triple box, i.e. set $a_i=1$ and
apply, after a preliminary analysis, a standard procedure of
taking residues and shifting contours which results in
contributions where one can expand in $\ep$ in the integrand.
In this way, it is not immediately clear how to perform integrations
in some of 2-dimensional MB integrals obtained. One can however proceed
numerically at this stage and arrive at a result in expansion in $\ep$
where the simple pole and the finite part are numerically evaluated.

On the other hand, one can follow a straightforward method based
on the MB representation and organize the evaluation of the leading
power asymptotic behaviour
in such a way that it will be a part of a future evaluation of the
unexpanded triple box.
Using the same way of introducing MB integrations as outlined above,
after appropriate changes of variables,
we arrive at the following (only!) sevenfold MB representation of
the (unexpanded) triple box (\ref{3box}):
\bea
T(a_1,\ldots,a_8;s,t ;\ep)
 &=&
\frac{\left(i\pi^{d/2} \right)^3 (-1)^a}{
\prod_{j=2,5,7,8,9,10}\Gm(a_j) \Gm(4-a_{589(10)}-2\ep)(-s)^{a-6+3\ep}}
\nn \\ &&  \hspace*{-50mm}\times
\frac{1}{(2\pi i)^7} \int_{-i\infty}^{+i\infty}
\dd w \prod_{j=2}^7 \dd z_j
\left(\frac{t}{s} \right)^{w}
\frac{\Gm(a_{2} + w)\Gm(-w)
\Gm(z_2 + z_4) \Gm(z_3 + z_4)}
{\Gm(a_1 + z_3 + z_4) \Gm(a_3 + z_2 + z_4)}
\nn \\ &&  \hspace*{-50mm}\times
\frac{
\Gm(2 - a_{12} - \ep + z_{2})
\Gm(2 - a_{23} - \ep + z_{3})
 \Gm(a_{7} + w - z_{4})  \Gm(-z_{5}) \Gm(-z_{6})
  }
{ \Gm(4 - a_{123} - 2 \ep + w - z_{4}) \Gm(a_{6} - z_{5})
  \Gm(a_{4} - z_{6}) }
\nn \\ &&  \hspace*{-50mm}\times
\Gm( + a_{123}-2 + \ep + z_{4})
 \Gm(w + z_{2} + z_{3} + z_{4} - z_{7})
\nn \\ &&  \hspace*{-50mm}\times
\Gm(2  - a_{59(10)} - \ep - z_{5} - z_{7})
  \Gm(2 - a_{589} - \ep - z_{6} - z_{7})
\nn \\ &&  \hspace*{-50mm}\times
  \Gm(a_{467}-2  + \ep + w - z_{4} - z_{5} - z_{6} - z_{7})
  \Gm(a_{9} + z_{7}) \Gm(a_{5} + z_{5} + z_{6} + z_{7})
\nn \\ &&  \hspace*{-50mm}\times
\Gm(4 - a_{467} - 2 \ep + z_{5} + z_{6} + z_{7})
 \Gm(a_{589(10)}-2+ \ep + z_{5} + z_{6} + z_{7})
\nn \\ &&  \hspace*{-50mm}\times
 \Gm(2 - a_{67} - \ep - w - z_{2} + z_{5} +
    z_{7}) \Gm(2 - a_{47} - \ep - w - z_{3} + z_{6} + z_{7})
\, .
\label{7MB}
\eea

In the case of the master triple box, we set $a_i=1$ for
$i=1,2,\ldots,10$ and obtain
\bea
T^{(0)}(s,t;\ep)\equiv T(1,\ldots,1;s,t;\ep) &&
\nn \\ &&  \hspace*{-75mm}
= \frac{\left(i\pi^{d/2} \right)^3}{
\Gm(-2\ep)(-s)^{4+3\ep}}
\frac{1}{(2\pi i)^7} \int_{-i\infty}^{+i\infty}
\dd w \prod_{j=2}^7 \dd z_j
\left(\frac{t}{s} \right)^{w}
\frac{ \Gm(1 + w)\Gm(-w) }{\Gm(1 - 2 \ep + w - z_4) }
\nn \\ &&  \hspace*{-75mm}\times
\frac{ \Gm(-\ep + z_2)
  \Gm(-\ep + z_3) \Gm(1 + w - z_4) \Gm(-z_2 - z_3 - z_4)
  \Gm(1 + \ep + z_4)}
{ \Gm(1 + z_2 + z_4)
  \Gm(1 + z_3 + z_4)}
\nn \\ &&  \hspace*{-75mm}\times
\frac{  \Gm(z_2 + z_4) \Gm(z_3 + z_4) \Gm(-z_5)
  \Gm(-z_6) \Gm(w + z_2 + z_3 + z_4 - z_7) }
{ \Gm(1 - z_5) \Gm(1 - z_6) \Gm(1 - 2 \ep + z_5 + z_6 + z_7) }
\nn \\ &&  \hspace*{-75mm}\times
 \Gm(-1 - \ep - z_5 - z_7) \Gm(-1 - \ep - z_6 - z_7)\Gm(1 + z_7)
\nn \\ &&  \hspace*{-75mm}\times
  \Gm(1 + \ep + w - z_4 - z_5 - z_6 - z_7)
  \Gm(-\ep - w - z_2 + z_5 + z_7)
\nn \\ &&  \hspace*{-75mm}\times
 \Gm(-\ep - w - z_3 + z_6 + z_7)
  \Gm(1 + z_5 + z_6 + z_7) \Gm(2 + \ep + z_5 + z_6 + z_7)
\, .
\label{7MB0}
\eea
Observe that, because of the presence of the factor $\Gm(-2\ep)$
in the denominator, we are forced to take some residue
in order to arrive at a non-zero result at $\ep=0$,
so that the integral is effectively sixfold.

The asymptotic expansion in the Regge limit is determined, in the
language of the MB representation (\ref{7MB0}), by the poles in
the variable $w$
that come from a $-w$ dependence of the gamma functions.
The poles of $\Gm(-w)$ correspond to the hard contribution to the
asymptotic expansion (within expansion by regions) which turns out to
be subleading (order $t^0$). The rest of the poles are not explicit
and are generated due to integration over the other variables, $z_i$.
An analysis of the integrand shows that the leading Regge
asymptotic behaviour,
which is effectively described by a gamma function of the
type $\Gm(-1-3\ep-w)$,
is generated by the gamma functions
\[\Gm(-\ep + z_2) \Gm(-\ep - w - z_2 + z_6 + z_7)
\Gm(-1 - \ep - z_6 - z_7)\]
or, symmetrically, by
\[\Gm(-\ep + z_3) \Gm(-\ep - w - z_3 + z_5 + z_7)
\Gm(-1 - \ep - z_5 - z_7).\]
Then the standard procedure of shifting contours and taking residues
is applied. It results again in a sum of MB integrals where a
Laurent expansion in $\ep$ in the integrand becomes possible.

{}Following this second variant of evaluation we see that all
integration over $z_i$ variables are then explicitly performed
by means of the first and the second Barnes lemmas and their corollaries.
The last integral, over $w$, is taken by closing the integration
contour in the $w$ complex plane to the right and taking a series
of residues at $w=0,1,2,\ldots$
At the final stage, summation formulae for series involving $1/n^j$,
$S_k(n)$, $S_{ik}(n) =  \sum_{j=1}^n j^{-i} S_{k}(j)\,,$ etc.
are used. Some of them are presented in Appendix.

The final result for the Regge asymptotics of the planar
triple box takes the following form:
\bea
T^{(0)}(s,t;\ep) &=&
-\frac{\left(i\pi^{d/2}
\e^{-\gm_{\rm E}\ep} \right)^3}{s^3 (-t)^{1+3\ep}}
\left\{
\frac{16}{9\ep^6}-\frac{5 L}{3 \ep^5}-\frac{3 \pi^2}{2 \ep^4}
- \left[\frac{11\pi^2}{12} L  + \frac{131 \zeta(3)}{9}
\right]\frac{1}{\ep^3}
\right.
\nn \\ &&  \hspace*{-10mm}
+  \left[\frac{49 \zeta(3)}{3} L- \frac{1411\pi^4}{1080}\right]
\frac{1}{\ep^2}
-\left[\frac{503\pi^4 }{1440 }L  - \frac{73 \pi^2\zeta(3)}{4}
+ \frac{301 \zeta(5)}{15}\right]\frac{1}{\ep}
\nn \\ &&  \hspace*{-10mm} \left.
+\left[\frac{223 \pi^2 \zeta(3)}{12}  + 149 \zeta(5)\right] L
-\frac{624607 \pi^6 }{544320}  + \frac{167 \zeta(3)^2 }{9}
+ O(\ep) \right\}
\label{Result} \;,
\eea
where $L=\ln s/t$ and $\zeta(z)$ is the Riemann zeta function.

A non-trivial check of this result comes from the first variant
of the evaluation of the Regge asymptotic presented above where
the coefficients at the poles $1/\ep^j,\; j=2,\ldots,6$ agree
exactly with (\ref{Result}) while the coefficient at $1/\ep$
and the finite part in $\ep$ agree numerically.

The value $16/9$ for the coefficient at the highest pole is also
in agreement with the explicit result for a general L-loop ladder planar
diagram\footnote{This result stays unique, starting from
the two-loop level, for the class of massless four-point diagrams with
all legs off-shell. For example, if one contracts some line
(other than a rung) or puts
a dot on some line, no analytical results are available for the moment.}
(L-box) by Davydychev and Ussyukina \cite{UD} if we believe in
a well-known empirical rule which corresponds
$1/\ep$ to a logarithm of an expansion parameter. From their
result one can find the leading power and logarithmic
asymptotics:
$(i \pi^2)^L 4^L /(L! s^L t) \ln^{2 L}(-s).$
This gives, in particular, $4$ for $L=1$ and $L=2$, and $16/9$ for  $L=3$,
in agreement with existing results at one-, two- and (now) at
three-loop level. For higher poles, this correspondence
hardly exists, even in some generalized form.

The coefficients at $1/\ep^6$,..., $1/\ep^3$ in (\ref{Result}) have
been also confirmed by a numerical check \cite{BH1}
with the help of a method  \cite{BH} based on numerical integration in
the space of alpha parameters.

The procedure described above can be applied, in a similar way, to the
calculation of Regge asymptotics of any massless planar on-shell
triple box.

\vspace{0.5 cm}

{\em Acknowledgments.}
I am grateful to V.S.~Fadin and A.A.~Penin for helpful discussions
of perspectives of the three-loop BFKL analysis.
Thanks again to G.~Heinrich for
the numerical check of the highest poles.
This work was supported by the Russian Foundation for Basic
Research through project 01-02-16171, INTAS through grant 00-00313,
and the Volkswagen Foundation, contract No.~I/77788.

\appendix
\section{Summation Formulae}
Summation formulae with at least $1/n^2$ factor,
which explicitly provides convergence, for example,
\[
\sum_{n=1}^{\infty} S_1(n-1)  S_{12}(n-1) \frac{1}{n^2} =
\frac{313 \pi^6}{45360}  - 2 \zeta(3)^2\, ,
\]
can be obtained\footnote{
I have derived these formulae myself and then was informed about
existence of these procedures. For a person who has never used FORM,
it looks simpler to derive (and check) these formulae
than to learn how to run FORM and use SUMMER (and check results
obtained.)}
using FORM \cite{FORM} procedures called SUMMER and described
in \cite{Harm}.

Here is a table of summation formulae with the factor $1/n$.
They are not explicitly present in SUMMER.
The convergence is here provided by other factors,
$ \psi^{(k)}(n+1)= (-1)^k k! \left[
S_{k+1}(n)-\zeta(k+1)\right],\; k=1,2,\ldots$.
We have
\bea
\sum_{n=1}^{\infty}  \psi''''(n+1)\frac{1}{n}  &=&  -\frac{2 \pi^6}{105}
+ 12 \zeta(3)^2\, , \\
\sum_{n=1}^{\infty}  \psi'''(n+1) S_1(n) \frac{1}{n}  &=&
\frac{\pi^6}{1512} \, ,
\\
\sum_{n=1}^{\infty}  \psi''(n+1) S_1(n)^2 \frac{1}{n}  &=&
\frac{\pi^6}{90}
- 8 \zeta(3)^2\, ,
\\
\sum_{n=1}^{\infty}  \psi'(n+1)^2 S_1(n)\frac{1}{n}  &=&
-\frac{\pi^6}{432}
+ 2 \zeta(3)^2\, ,
\\
\sum_{n=1}^{\infty}  \psi'(n+1) S_1(n)^3\frac{1}{n}  &=&
\frac{269 \pi^6}{22680} \, ,
\\
\sum_{n=1}^{\infty}  \psi'(n+1) \psi''(n+1) \frac{1}{n}  &=&
\frac{61 \pi^6}{22680}  - 2 \zeta(3)^2 \, ,
\\
\sum_{n=1}^{\infty}  \psi'''(n+1) \frac{1}{n}  &=&
- \pi^2 \zeta(3)  + 12 \zeta(5)\, ,
\\
\sum_{n=1}^{\infty}  \psi'(n+1) S_1(n)^2\frac{1}{n}  &=&
\frac{\pi^2 \zeta(3)}{3} \, ,
\\
\sum_{n=1}^{\infty}  \psi''(n+1) S_1(n)\frac{1}{n}  &=&
-\frac{2 \pi^2 \zeta(3)}{3}  +  7 \zeta(5)\, ,
\\
\sum_{n=1}^{\infty} \psi'(n+1)^2 \frac{1}{n} &=&
\frac{5 \pi^2 \zeta(3)}{6} - 9 \zeta(5) \, ,
\\
\sum_{n=1}^{\infty} \psi''(n+1)\frac{1}{n}  &=&  -\frac{\pi^4}{180} \, ,
\\
\sum_{n=1}^{\infty}  \psi'(n+1) S_1(n)\frac{1}{n}  &=&
\frac{7 \pi^4}{360} \, ,
\\
\sum_{n=1}^{\infty}  \psi'(n+1)\frac{1}{n}  &=&  \zeta(3) \, .
\label{sumNum1}
\eea

\end{document}